\documentclass[letterpaper]{sig-alternate-2013}

\makeatletter
\def\@copyrightspace{\relax}
\makeatother

\usepackage[T1]{fontenc}

\usepackage{xcolor}
\usepackage{amsmath}
\usepackage{balance}
\usepackage{verbatim}
\usepackage{graphics}
\usepackage{subfig}
\usepackage{hyperref}
\usepackage{times}

\newtheorem{pty}{Property}

\newcommand\reffig[1]{Fig.~\ref{#1}}
\newcommand\tref[1]{Table~\ref{#1}}
\newcommand\tabhead[1]{\small\textbf{#1}}

\title{Social Bootstrapping: How Pinterest and Last.fm Social Communities Benefit by Borrowing Links from Facebook\\ {\LARGE [Please cite the WWW'14 version of this paper]}}

\numberofauthors{1}
\author{Changtao Zhong$^1$, Mostafa Salehi$^2$, Sunil Shah$^3$, Marius Cobzarenco$^4$, \\Nishanth Sastry$^1$,  Meeyoung Cha$^5$\\
\affaddr{$^1$King's College London  $^2$University of Tehran $^3$UC Berkeley $^4$Last.fm $^5$KAIST}\\
\affaddr{$^1$\{changtao.zhong, nishanth.sastry\}@kcl.ac.uk, $^2$mostafa\_salehi@ut.ac.ir, $^3$sunil.shah@berkeley.edu,}\\
\affaddr{$^4$marius@last.fm, $^5$meeyoungcha@kaist.edu}
}

\begin{document}
\maketitle
\begin{abstract}
How does one develop a new online community that is highly engaging to each user and promotes social interaction? A number of websites offer friend-finding features that help users bootstrap social networks on the website by copying links from an established network like Facebook or Twitter. This paper quantifies the extent to which such \emph{social bootstrapping} is effective in enhancing a social experience of the website. First, we develop a stylised analytical model that suggests that copying  tends to produce a giant connected component (i.e., a connected community) quickly and preserves properties such as reciprocity and clustering, up to a linear multiplicative factor. Second, we use data from two websites, Pinterest and Last.fm, to empirically compare the subgraph of links copied from Facebook to links created natively. We find that the copied subgraph has a giant component, higher reciprocity and clustering, and confirm that the copied connections see higher social interactions. However, the need for copying diminishes as users become more active and influential. Such users tend to create links natively on the website, to users who are more similar to them than their Facebook friends. Our findings give new insights into understanding how bootstrapping from established social networks can help engage new users by enhancing social interactivity.
\end{abstract}

 \category{H.3.5}{Online Information Services}{Commercial Services, Data Sharing, Web-based services}
 \category{J.4}{Social and Behavioral Sciences}{Sociology}

 \keywords{Social Bootstrapping; Friend Finder Tools; Community Design; Social Property; Social Interaction; Copied Networks}

\section{Introduction}
How to design online communities and maintain users participation is a fundamental problem for website designers. 
Many  websites now try to incorporate a social networking aspect to enhance user engagement and create active communities. Making a website ``social'' typically involves linking users together and providing some kind of awareness of the linked users' activities to each other. Studies have found that such social networking aspects facilitate community formation in learning ~\cite{baird_neomillennial_2005,conole_design_2010,heiberger_have_2008}, working~\cite{dimicco_motivations_2008,lee_experiments_2013}, medicine \cite{eysenbach_medicine_2008} and online games~\cite{choi_why_2004,ducheneaut_social_2004} applications. 

However, in creating such a social experience on a website, designers face an important choice: should they create an entirely new social network embedded within the site? Or should they instead connect users who are already linked together on an established social network such as Facebook and Twitter? The latter option has recently become a possibility, with both Facebook and Twitter opening up their social graphs to third-party websites, who can write friend-finder tools that help users select and import friendship links from these established networks into their own service (e.g., through the open graph protocol~\cite{fb-opengraph}). 

We term this act of copying existing friends from an established social network onto a third-party website as \textit{social bootstrapping}.  
Social bootstrapping has direct implications on how a new online social network community can grow quickly. However, this problem is complex to examine with real data because it involves user interaction across  multiple  heterogeneous networks. To this end, we gather massive amounts of data from Facebook, Pinterest, and Last.fm involving tens of millions of nodes and billions of links and explore the potential benefits and limitations of social bootstrapping\footnote{The Pinterest datasets used in this paper are shared for wider community use at \url{http://www.inf.kcl.ac.uk/staff/nrs/projects/cd-gain/social_bootstrapping.html}. }. We seek to evaluate how such bootstrapping could affect the user community and to what extent copying links contributes to social structure and user engagement as the new website matures. 

Although copying clearly enriches the number of social links on the new website, it is not \textit{a priori} clear whether links borrowed from a general-purpose social network such as Facebook would be appropriate for content-driven sites, which typically attempt to link users  interested in similar content. Additionally, social bootstrapping involves a two-step process. First, to copy a Facebook friendship, for example, both users joined by the friend link have to independently decide to connect their accounts on the third-party website with their Facebook accounts.  Next, they have to select which of their friends to import into the third-party website and choose this particular link. Thus, social bootstrapping can be limited in the number of links that get copied over.

Nonetheless, social bootstrapping becomes effective for user engagement and community formation if it creates the sort of structure conducive to social interaction and increased user activity on the third-party website. Therefore, using a combination of analytical models and empirical studies, we focus on  three important structural properties, namely connectivity, reciprocity, and clustering. We study the effect of copying on these properties and how social interactions are affected in turn. We also study how copying evolves among more active and influential users and build up a picture of the importance of creating native links on the new website.

We first develop a stylized model of copying as a process of sampling links from the established network. To mimic the two-step filtering process described above, we propose the Link Bootstrapping Sampling (LBS) model as a variation of  induced subgraph  sampling~\cite{Kolaczyk_network_2009}. Under this simple analytical model, 
  we study the emergence of a giant connected component in the copied network. We demonstrate that when copying from a typical network with a heavy-tailed degree distribution, a giant component emerges even with a small amount of sampling, which suggests that social bootstrapping may be an effective means of increasing user engagement and creating a  connected community.

We then empirically study copying, using data from two large websites, Pinterest and Last.fm, which include friend-finding tools to copy friends from Facebook. We make efforts to tease out various social effects. We  study the structural properties of the copied subgraph, comparing it to the subgraph of links created natively on the website, and find that copying enriches reciprocity and clustering of the local structure. Both reciprocity and clustering are shown to be important for social interactions, indicating that social bootstrapping successfully promotes user engagement. 

However, copying links yields diminishing returns. As users become more active and influential on the new website, they create proportionally more native links than copied ones. Native links offer a benefit over copied links: users connecting natively on Pinterest and Last.fm tend to be more similar to each other in their tastes than with the ones copied from Facebook. This is an important observation for long-term user engagement, as prolific users tend to engage more with native links and fine-tune the local relationships to meet their interests.  As a result, we conclude that while ``copying'' links is essential to bootstrap one's network, the opposite  ``weaning''  process is equally important for long lasting user engagement. 

To the best of our knowledge, this paper is the first to demonstrate how content-driven websites like Pinterest and Last.fm can benefit from social bootstrapping by copying links from established networks like Facebook. Through extensive analysis of cross-net\-work data, we are able to describe both how new social communities are seeded by social bootstrapping, and how users grow beyond the bootstrapped links to create strong communities natively. We believe our findings have strong implications for the design of new content-driven web communities. 


\section{A Link Bootstrapping Model}
\label{sec:analysis}
In this section, we propose a simple analytical model of social bootstrapping to gain insight about its implications on network structure. Our model allows us to analytically examine how copying links affects key structural features that facilitate social interactions in the target network, such as reciprocity, clustering and the formation of a Giant Connected Component (GCC).
\subsection{Terminology}
\label{sec:terminology}
\noindent

Social bootstrapping refers to the act of copying existing friend links from a source social network onto a third-party website to create a target social network. We define several sub-networks below to describe this phenomenon:

\begin{figure}[tp]
\centering
\includegraphics[width=0.9\columnwidth]{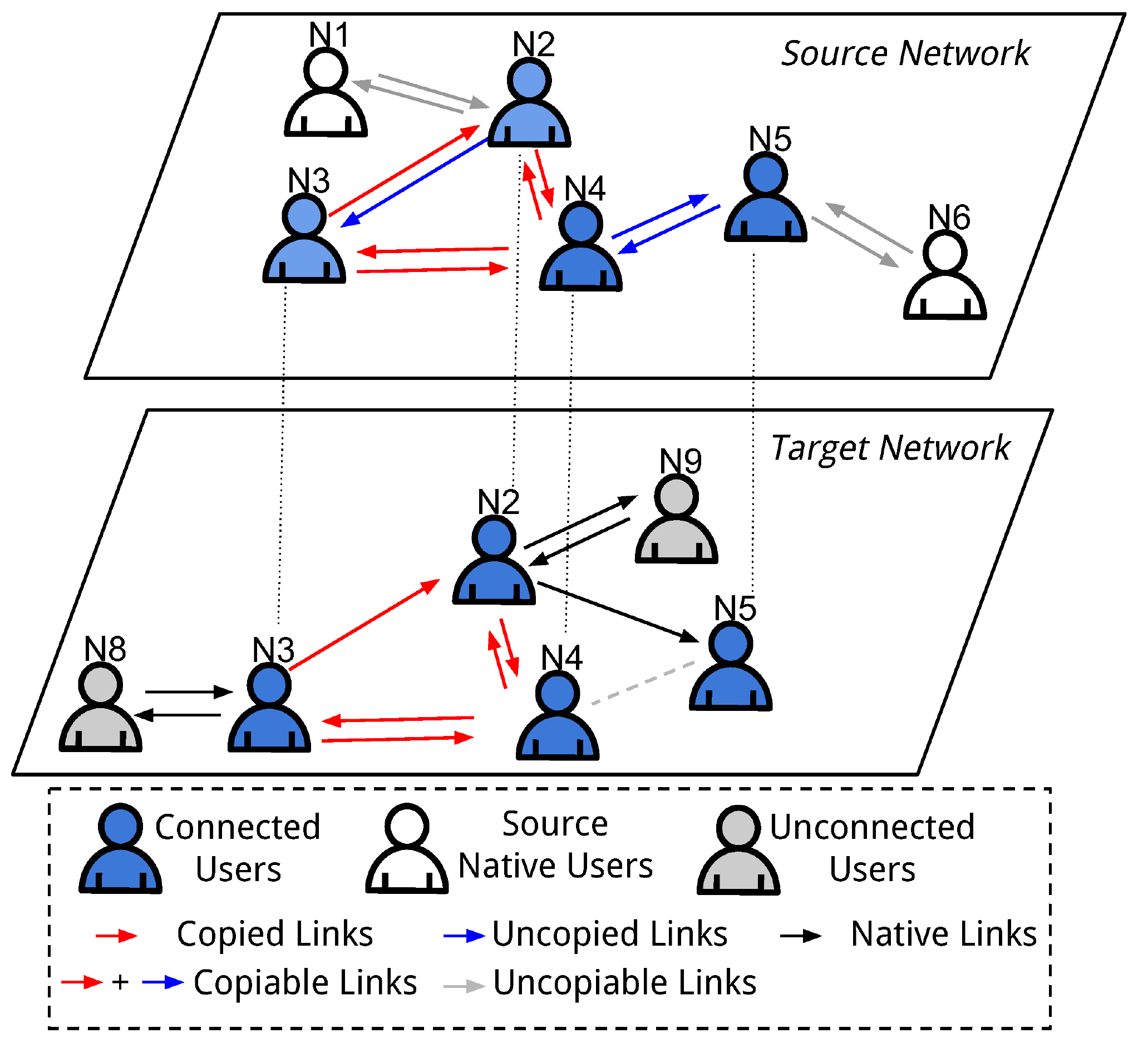}
\caption{The structure of social bootstrapping.}
\label{fig:subgraph}
\vspace{-4mm}
\end{figure}

\begin{description}
\setlength{\itemsep}{0pt}
    \setlength{\parsep}{0pt}
    \setlength{\topsep}{0pt}
    \setlength{\partopsep}{0pt}
    \setlength{\leftmargin}{2em}
    \setlength{\labelsep}{0.5em}
\item[Source network:] The social graph of an established social network like Facebook (Fb for short), which contains a significant number of nodes and links (e.g., 1.19 billion monthly active users as of 2014\footnote{\url{http://newsroom.fb.com/Key-Facts}}). The source network is displayed as the upper layer in the toy example in \reffig{fig:subgraph}.  Note that some users, such as $N1$ and $N6$ are \emph{source native} and are present only in the source network.

\item[Target network:] The relatively new third-party network that allows users to copy links from established networks, displayed as the lower layer in \reffig{fig:subgraph}. \emph{Connected nodes} are the subset of all nodes in target network that have used the ``Friend Finder'' tool to connect their accounts to the source network. In the toy example, blue nodes, i.e., $N2$, $N3$, $N4$ and $N5$ are  the connected nodes. Grey nodes, i.e., $N8$, $N9$ are \emph{unconnected nodes} who either exist only on the target network or have chosen not to connect their accounts on the source network to their identity on the target network. Within the target network, social links copied from the source network are called  \emph{copied links} and those created natively are called  \emph{native links}. Copied links in the target network may be directed even if they are copied from the undirected source network. Copied links are a subset of \emph{copiable} links, the set of all links between connected nodes in the source network. We take Pinterest (\emph{Pnt} for short) and Last.fm (\emph{Lfm} for short) as two target networks of interest. 

\item[Copied network:] The social subgraph of the target network solely containing copied links and all \emph{connected nodes}. In \reffig{fig:subgraph}, the copied network contains the red edges and all blue nodes. We call the  network copied from Facebook as \emph{Fb-copied}.

\item[Native network:] The subgraph of the target network that only contains native links and the corresponding nodes at either end of each native link. In the toy example, the native network is the subgraph made up by black edges and nodes linked by them, i.e., $N2$, $N3$, $N5$, $N8$ and $N9$. Nodes can be in copied and native networks at the same time, but links are either copied or native. We call the native networks for Pinterest and Last.fm \emph{pnt-native} and \emph{lfm-native}, respectively.

\end{description}

\subsection{A random bootstrapping process}
We now introduce a simple analytical model that represents copying a social link as a simplified random sampling process. We propose a two-step model, called the \textit{Link Bootstrapping Sampling (LBS)}, which is a variation of the induced subgraph sampling process~\cite{Kolaczyk_network_2009}. In the first step, users of the target network have to self-select to connect their accounts on the target network with the source network. In the second step, users have to select which of their friends from the source network to import onto the target network. Under this stylized model, we obtain expressions for the resulting degree distributions of the copied network and a condition for the emergence of a giant connected component in that network. Although the model considers  directed target networks, it can be trivially adapted for undirected networks.

Formally, let $G = (V, E)$ be the graph representing the globally-established source network, where $V$ is the set of nodes, and $E$ is the set of links between pairs of nodes. We assume that the LBS copying process randomly samples each node $N_i$ with a probability $p_1(N_i)$. We call this the \emph{node sampling rate}. Each selected node independently selects each of its neighbours $N_j$ in the original network with a probability $p_2(N_i,N_j)$. This is the \emph{link sampling rate}. 
Let $S \subset V$ be a sample of nodes, and $L \subset E$ denote the collection of sampled links to be found in $G$ among subset $S$. Then, the subgraph $G(S) = (S,L)$ represents a copied network. A (directed) copied link from $N_i$  to $N_j$ appears in $G(S) = (S,L)$ if both events (node and link sampling) happen, i.e., with probability $p_1(\cdot) p_2(\cdot,\cdot)$. We call this the \textit{link copy probability}. 

\subsection{Giant component in the copied subgraph}
We assume for ease of exposition that the node and link sampling rates $p_1$ and $p_2$ are uniform across all nodes and edges respectively to set the following link copy probability
\begin{equation}
p_e = p_1 p_2
\end{equation}
 However, similar results on the appearance of GCC, Eq.~\eqref{GCCappear}, can also be obtained using mean field approximations for $p_e$. This allows us to consider more realistic assumptions such as the node sampling rate being proportional to the degree to reflect the possibility that more social nodes connect their source and target accounts with higher probability. Alternatively, we may also consider conditions such as the link sampling rate being proportional to the  number of common friends between the nodes on either side of the link to reflect the possibility that socially closer links are copied over with higher probability.  

The probability that a node 
will have exactly $k^{i(o)}$ links in the copied subgraph $G(S)$ is given by: 
\begin{multline}
\label{sampledDegreeDist}
p_c^{i(o)}{(k^{i(o)})} = \sum_{k^{i(o)}_0=k^{i(o)}}^{\infty} p_s{(k^{i(o)}_0)}
\\
\times \dbinom{k^{i(o)}_0}{k^{i(o)}} {(p_e)}^{k^{i(o)}} {(1 - p_e)}^{k^{i(o)}_0 - k^{i(o)}}
\end{multline}
where $p_s^{i(o)}(k)$ is the in- (out-) degree distribution of the source network $G$.
Similarly, the joint degree distribution of obtaining a node with in-degree $j$ and out-degree $k$ in the copied subgraph is:
\begin{multline}
\label{sampledJointDegreeDist}
p_c(j,k) = \sum_{j_0 = j}^{\infty}\sum_{k_0 = k}^{\infty} p_s(j_0,k_0) \dbinom{j_0}{j} {(p_e)}^{j} {(1 - p_e)}^{j_0-j}
\\
\times \dbinom{k_0}{k} {(p_e)}^{k} {(1 - p_e)}^{k_0-k}
\end{multline} 
where $p_s{(j,k)}$ is the joint degree distribution of source network $G$.

We are now ready to state an important property, relating the link copy probability to the emergence of a giant component: 
\begin{pty}
When copying from an undirected source network (or equivalently, every link occurs in both directions), a giant component appears in the copied network if
\begin{equation}
\label{GCCappear}
p_e \geq {{{{\langle k \rangle}}^{\prime}} \over {{\langle k^2 \rangle}^{\prime}}},
\end{equation}
where averages are computed with respect to the degree distribution in the source network.
\end{pty}

\begin{proof}
In a directed  network of arbitrary degree distribution $p(j,k)$, a GCC exists if \cite{schwartz_2002}:
\begin{equation}
\label{gccDirCriteria}
{{\langle jk \rangle}} \geq {{\langle k \rangle}}
\end{equation}

For the copied network induced by the Link Bootstrapping Sampling, the average of the degree and joint degree distribution of the subgraph $G(S)$ sampled by the Link Bootstrapping Sampling method from a network $G$ with joint distribution  $p_s{(j,k)}$, can be calculated, respectively, as
$\langle jk \rangle$ = $p_e^2 \langle jk \rangle^\prime$ and $\langle k \rangle$ = $p_e \langle k \rangle^\prime$.
where ${{\langle jk \rangle}}^{\prime}$ and ${{\langle k \rangle}}^{\prime}$ are computed with respect to the original joint degree distribution (i.e., $p_s{(j,k)}$).
Thus, we can rewrite Eq.~\eqref{gccDirCriteria}  as:  
\begin{equation}
\label{samplingGCC}
p_e \geq {{{\langle k \rangle}^{\prime}} \over {{\langle jk \rangle}}^{\prime}}
\end{equation}
Since the source network is undirected, in- and out-degrees are completely correlated. i.e.,  ${{\langle jk \rangle}}^{\prime}={{\langle k^2 \rangle}}^{\prime}$. With this, Eq.~(\ref{samplingGCC}) reduces to Eq.~\eqref{GCCappear}.
\end{proof}

Thus the link copy probability $p_e$ at which a GCC emerges in the sampled subgraph under Link Bootstrapping Sampling depends in a simple and intuitive manner on the properties of the source network.
For instance, if the source network can be modeled as an Erdos-R\'enyi random network, where the degree distribution is given by a Poisson distribution with parameter $\lambda$, i.e. $p_s(k)= {{{\lambda}^k exp(-{\lambda})} \over {k!}}$, the first and second moments are given by ${\langle k \rangle}^{\prime}=\lambda$ and ${\langle k^2 \rangle}^{\prime}={\lambda}^2+\lambda$. Thus, the link copy probability must exceed $p_e > {{1} \over {\lambda+1}}$ for a GCC to exist. If the source network is scale-free, a GCC emerges easily if the degree distribution of the source network has infinite second moments ${\langle k^2 \rangle}^{\prime}$. In particular, we know that in a power law distribution $p_s(k)= c{k}^{-\gamma_s}$, all moments of order $m > \gamma_s - 1$ are infinite. Thus, if the Link Bootstrapping Sampling is applied to copy links from an undirected \emph{scale-free network} with power law exponent $\gamma_s < 3$, a GCC will come into existence even with very low link copy probability ($p_e \rightarrow 0$). In general, the larger the second moment, or equivalently, the larger the variation in the degree distribution, the easier it is (i.e., the lower the link copy probability needed) for a giant component to emerge.

\subsection{Other properties}
Next, we study the effect of Link Bootstrapping Sampling (using uniform link and node sampling rates) on two other properties in the copied subgraph, which are thought to be correlated with social interaction: reciprocity and clustering coefficient. Both increase proportionally with the link sampling rate.
\subsubsection{Reciprocity}
\label{sec:reciprocity}
First, we study  the effect of bootstrapping on $R^c$, the reciprocity of the copied network, defined as the proportion of links which exist in both directions, among all copied links.  
We have 

\begin{equation}
\label{rec_c}
R^c= 1- {{2 m_s [p_2 (1-p_2)]}\over{2 m_s p_2}}= p_2
\end{equation}
where $m_s$ is the number of links in the source network. Thus, reciprocity is defined by the link sampling rate; higher link sampling rates result in higher expected reciprocity.
\subsubsection{Clustering}
\label{sec:clustering}

Next, we obtain an expression for the clustering coefficient of the copied network.  
Taking the copied network to be an uncorrelated undirected network with arbitrary degree distribution, the clustering coefficient takes the value \cite{Serrano2006}: 
\begin{equation}
\label{ccEquLBScopied}
C ^c= {1 \over n} {{[{\langle k^2 \rangle}-{\langle k \rangle}]^2} \over {{\langle k \rangle}^{3}}}
\end{equation}
where ${\langle k \rangle}$ and ${\langle k^2 \rangle}$ are the first and second moments of the degree distribution, respectively in the copied network, and $n$ is the number of sampled nodes. Writing these moments in terms of the corresponding moments ${\langle k \rangle}^{\prime}$, ${\langle k^2 \rangle}^\prime$ of the source network \cite{Son2012}:
\begin{equation}
\label{momentsSampled}
{\langle k \rangle} = {p_e} {\langle k \rangle}^{\prime}, \hspace{4mm}   
{\langle k^2 \rangle} = {p_e}^2 {\langle k^2 \rangle}^{\prime}+ {p_e} {(1-p_e)} {\langle k \rangle}^{\prime}
\end{equation}
Substituting these formulae into Eq.~(\ref{ccEquLBScopied}), we have 
\begin{multline}
\label{ccEquLBS2}
C ^c= {1 \over n} {{{p_e}^4[ {{\langle k^2 \rangle}^{\prime} - {\langle k \rangle}^{\prime}]^2} \over {{p_e}^3 {\langle k \rangle}^{^{\prime}3}}}}={1 \over n} p_e N C^s={p_1N \over n} {p_2} C^s= p_2C^s
\end{multline}
where $C^s$ is the clustering coefficient of the source network. 

The Link Bootstrapping Sampling with uniform node and edge sampling preserves the clustering coefficient of the source network, up to a multiplicative factor corresponding to the link sampling probability. This means that copying links from a source network that has a high level of clustering  results in a  copied network also with a proportionally high level of clustering.


\section{Datasets}
Having gained initial insight into copying from an analytical perspective, in the rest of the paper, we take an empirical approach and examine social bootstrapping using datasets from two very different websites, Pinterest and Last.fm. These are considered as target networks in our analysis. In both cases, we study copying from Facebook as the source network. Our datasets include extensive social graphs from both target websites, as well as corresponding graphs from Facebook. In addition, it includes nearly all activities from selected periods on both websites. {The data that we collected from Pinterest is shared to the research community, while much of the Last.fm data is already available through a public API.}

\subsection{Pinterest}

Pinterest is a photo sharing website that allows users to store and categorise images. Images added on Pinterest are termed \emph{pins} and can be created in two ways. The first way is \textit{pinning}, which imports from a URL external to pinterest.com. A second is \textit{repinning} from an existing pin. All pins are organised into \textit{pinboards} or {boards}, which belong to one of 32 globally specified {categories}. In addition to pinning or repinning, users can {like} a pin or {comment} on a pin.

The social graph of Pinterest is created through users \textit{following} other users or boards they find interesting. We call social links created in this way  {native links}. In addition to this method, users are able to connect with their Facebook and Twitter accounts and import their social links into Pinterest.  The \textit{Find Friends} function provides a list of Facebook and Twitter friends who are also registered on Pinterest. Users can select some of them to follow on the Pinterest website, which we call \textit{copied links}.

\tref{tab:datasets-pnt} summarizes the Pinterest dataset, consisting of the social graph on Pinterest, the corresponding nodes and edges on Facebook, and activities on the Pinterest site in January 2013. To obtain the Pinterest social graph, we used a snowball sampling technique, starting to crawl from a seed set of 1.6 million users which we collected in advance. In total 68.7 million Pinterest users and 3.8 billion directed edges between them were obtained. For each user, we checked whether there was a connected  Facebook account, and gathered basic profile information such as gender and profile, as well as basic statistics such as the number of pins, likes, followers, and followees. Of the 68.7 million, 40.4 million were Facebook-connected users, who have 2.4 billion links between them on Pinterest.

We next separate the 2.4 billion edges into those which are present on Facebook (i.e., are Fb-copied), and those which are native to Pinterest (Pnt-native).
To identify the Fb-copied portion of the network, we used the Facebook API to individually check whether a Pinterest link between two connected users was also present between the corresponding  Facebook accounts\footnote{Note that checking whether a pair of users are friends is affected by users' privacy setting. That is, it is unknown for us whether two users are friends or not if both of them had set their friend lists as private. Also, we assume that a link which exists both on Facebook and the target networks is a copied link, first made on Facebook and then copied  to the target network. Although we expect this to be the case normally, it is possible for user pairs to link to each other separately on Facebook and Pinterest, or link first on Pinterest, and subsequently on Facebook. We are unable to distinguish these cases from links copied using friend finder tools.}. We find that 0.98 billion links between connected users are also  on Facebook. These form our Fb-copied network. Pnt-native links were identified by excluding the Fb-copied network from our Pnt network.

 In a previous study~\cite{changtao_zhong_sharing_2013}, we had collected nearly all activities within Pinterest, during the period from January 3rd to 21st, 2013.  The crawl proceeded in two steps: first, to discover new pins, we visited each of the 32 category pages once every 5 minutes, and collected the latest pins of that category. Then, for every pin obtained this way, we visited the webpage of the pin every 10 minutes. A pin's webpage lists the 10 latest repins and the 24 latest likes\footnote{This setting has been changed in April 2013.}; we added these to our dataset, along with the approximate time of repins, likes and comments (if any). Through these regular visits, we captured almost all the activity during our observation period. We estimate that the fraction of visits which resulted in missed activities stands at $5.7 \times 10^{-6}$ for repins and $9.4 \times 10^{-7}$ for likes.
 In total, 8.5 million users (termed as \emph{active users}), 38.0 million repins and 19.9 million likes were included.

 Amongst these  active users, there are 5.2 million connected to Facebook. We crawled the Facebook pages of these 5.2 million connected active users, and attempted to obtain their Facebook friend lists. Due to  privacy settings, only 2.3 million users' social links could be obtained. Together, this collection of Facebook edges constitutes a subgraph of 444.2 million edges (\tref{tab:sampled-source-graph}). Of these, 141.9 million are \emph{copiable} links, i.e., edges between connected users who are on both Facebook and Pinterest.

\begin{table}[!htbp]
\vspace{-1mm}
  \centering
  \subfloat[Target social graph]{
  \begin{tabular}{|c|c|c|}
    \hline
    \tabhead{} & \tabhead{Nodes} & \tabhead{Links}\\
    \hline
	Pnt network  & 68,665,590 & 3,871,570,784\\
	Fb-copied & 40,472,339 & 983,520,986\\
    \hline
   \end{tabular}
   \label{tab:copy-pnt}
   }
\\
\vspace{-2mm}
  \subfloat[Activities in Pinterest]{
  \begin{tabular}{|c|c|c|c|}
    \hline
    \tabhead{} & \tabhead{Timespan} & \tabhead{Repins} & \tabhead{Likes}\\
    \hline
    Activities & 03-21 Jan 2013 & 38,041,368 & 19,907,874\\
    \hline
  \end{tabular}
  \label{tab:activity-pnt}
 }
\\
\vspace{-2mm}
 \subfloat[Facebook network]{
  \begin{tabular}{|c|c|}
    \hline
     	\tabhead{Nodes} & \tabhead{links}    \\
     	\hline
2,322,473 & 444,216,279   \\
    \hline
  \end{tabular}
  \label{tab:sampled-source-graph}
  }
\vspace{-3mm}
  \caption{The social graph among all Pinterest users.}
  \label{tab:datasets-pnt}
\vspace{-5mm}
\end{table}

\subsection{Last.fm}
Last.fm is a music discovery and recommendation website. Users can log what they listen to using a multitude of applications which support a variety of different operating systems and audio playback devices. This activity is known as \textit{scrobbling}. {Scrobbled} data is used to provide recommendations to users via collaborative filtering methods and is displayed publicly on users' profile pages. Users can \textit{love} tracks, a mechanism akin to {liking} a pin on Pinterest. These tracks are also displayed on their profile page and can optionally be shared to Facebook.

Last.fm offers a social network in which users can {friend} each other. A friendship between users can be considered as a bidirectional link, similar to that which Facebook offers. Friends of each user are displayed on their profile page, and when logged in, users are shown what their friends have scrobbled and what tracks their friends have loved.

Users can connect their Facebook accounts to Last.fm in three ways. The first is using their Facebook account to bootstrap basic profile information when they first sign up. Second, Last.fm offers a friend finder tool which connects to third party services such as Facebook, Google Mail and Yahoo! to look for contacts on those services, who also use Last.fm. Note that the  Facebook friend finder can only find other friends who have already connected their Facebook account to Last.fm. The third and most recent method is that users who share event attendance and loved tracks via their Facebook profile connect the two accounts as a result.

We considered a subset of the overall Last.fm user base by looking only at a sample of 1.8 million users who had, at some point in their history on Last.fm, connected a Facebook account to their Last.fm account using one of the methods. For each consenting user, we had access to their Last.fm social graph, basic profile information and their Facebook username. Of these users, 904,132 users use the Last.fm social features (i.e., have friendship edges in the Last.fm social network). Between these users, we extracted a subgraph of 12.3 million directed edges (or 6.15 million {friendships}) which forms the {Lfm} network.

For each of these 12.3 million Last.fm edges, we checked whether the friendship is also present in Facebook, using the Facebook API. Through this procedure, we identified 2.8 million copied edges between 600,000 users. Privacy settings meant that we were unable to validate friendships for approximately 200,000 users. We identify the {Lfm-native} network by eliminating copied edges.

We measure these users' activities on Last.fm in two ways: First, we measure listening activity   by counting their scrobbles. Second, we  measure  site usage from their website access log. Both measures cover the period Jan 1--Jun 22, 2013. 

Finally, we extract friend request data for requests sent during 2012 between Facebook connected users who were active on the site during that period (defined as those who have visited the site over 100 times during 2012). This data includes who initiated the friend request, as well as how it was made (i.e., through the friend finder tool, or natively on Last.fm) and whether the request was accepted, ignored, cancelled or is still pending. This  contains about 141,000 users and 1.1 million friend requests.

\vspace{-2mm}
\begin{table}[!htbp]
  \centering
  \begin{tabular}{|c|c|c|}
    \hline
    \tabhead{} & \tabhead{Nodes} & \tabhead{Links}\\
    \hline
Lfm network & 1,843,020 & 12,291,658 \\
Fb-copied & 592,992 & 2,787,000  \\
    \hline
  \end{tabular}
  \caption{Social graph of Last.fm users}
  \label{tab:datasets-active-users-lastfm}
  \vspace{-6mm}
\end{table}


\begin{figure*}[!htbp]
\centering
\hspace*{-5mm}
\subfloat[Reciprocity (copied vs native)]{
\includegraphics[width=0.72\columnwidth]{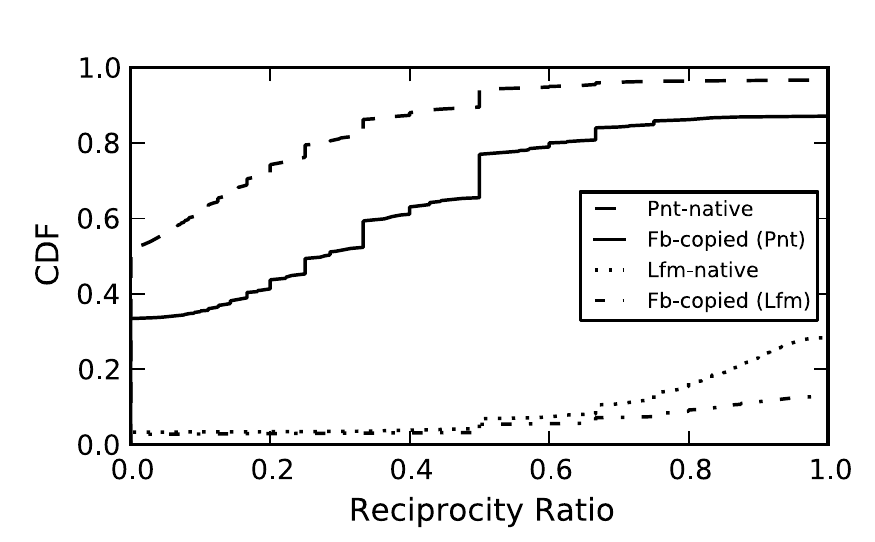}
\label{fig:reciprocity-cdf}
}
\subfloat[Clustering (copied vs native)]{
\includegraphics[width=0.72\columnwidth]{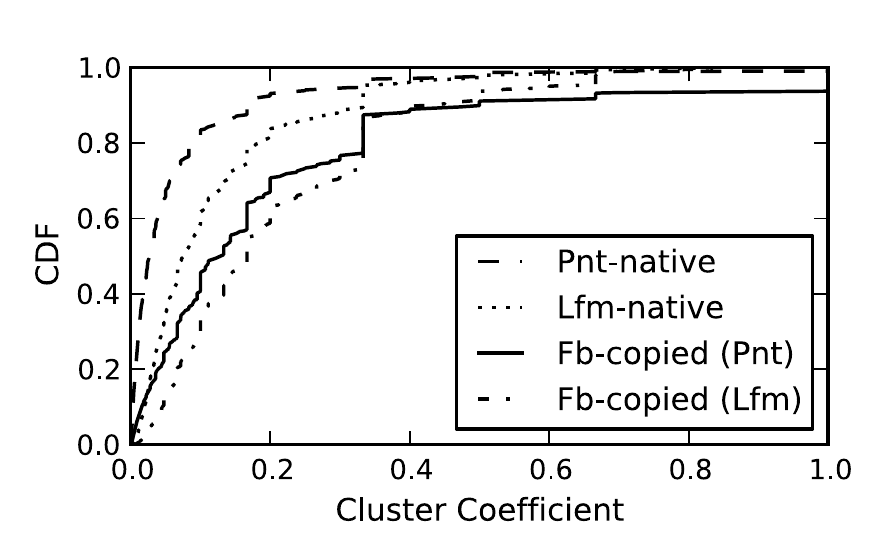}
\label{fig:cc-cdf}
}
\subfloat[Component sizes]{
\includegraphics[width=0.72\columnwidth]{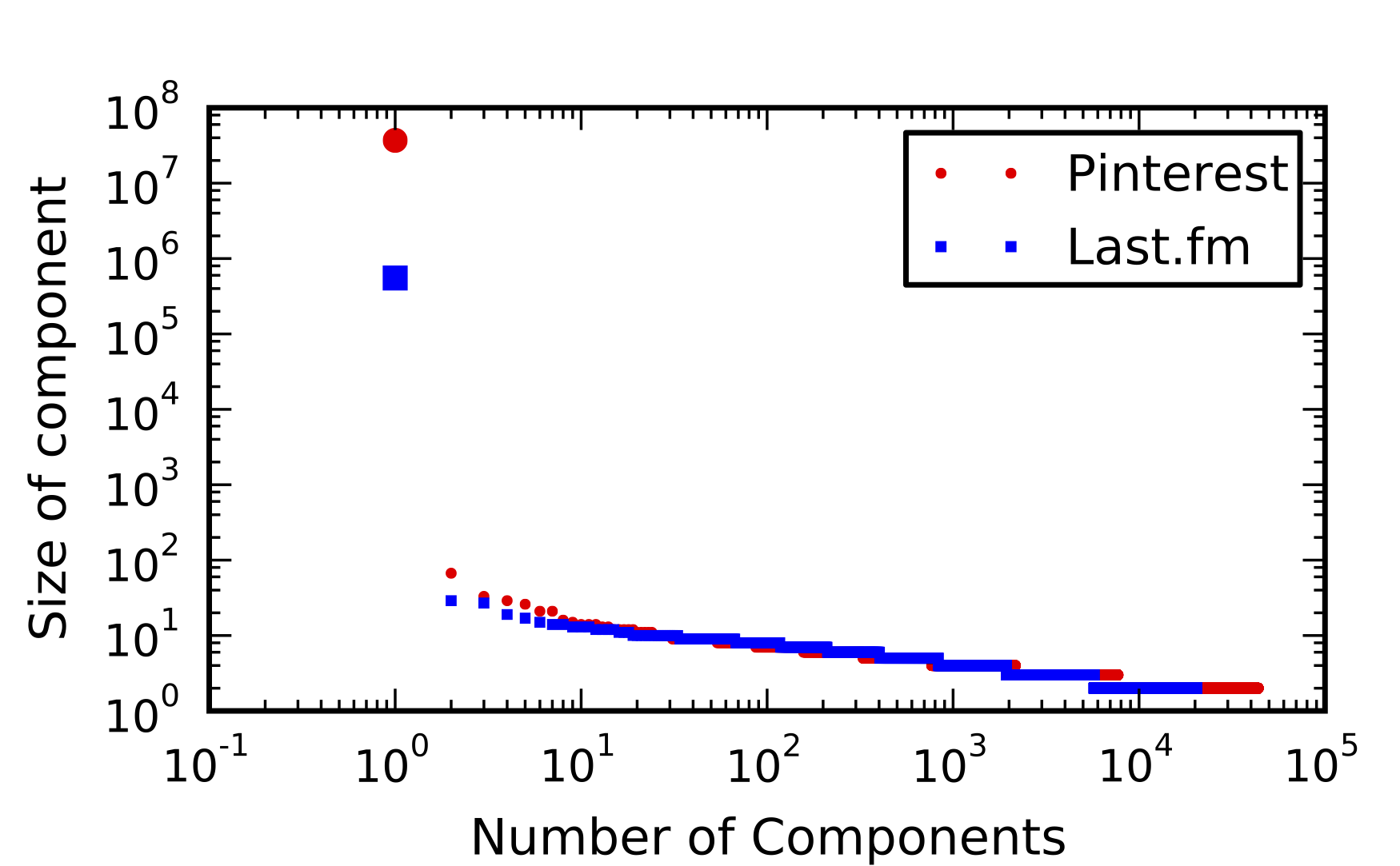}
\label{fig:components}
}
\caption{Properties of copied subgraph. (a) CDF of per-user fraction of links reciprocated in  copied and natively created networks. More links are reciprocated in the copied network. (b) Per-user CDF of clustering coefficients in natively created and copied subgraphs of Pinterest and Last.fm (0 valued-points not shown). Clustering coefficients are higher in the copied network. (c) Distribution of the sizes of connected components on the FB-Copied network in the Pinterest and Last.fm datasets. }
\end{figure*}

\section{Structural benefits of copying}
\label{sec:copy-properties}
Section~\ref{sec:analysis} showed that copying can produce desirable properties such as a giant component, reciprocity and clustering. We now empirically analyse how these structural properties of the copied network compare to the natively created network. We examine  implications of the differences we find, for social interaction in the target social network community. 

\subsection{Copied network has higher reciprocity}

Reciprocity is known to indicate positive bidirectional interaction between a pair of users, which is also known to increase user longevity in the system \cite{fehr2000far,boguna_generalized_2005,zhu_influence_2013}. Here, we attempt to examine the effect of copying on creating structurally stronger bidirectional social ties, by defining  \emph{reciprocity ratio} as the fraction of social links that are \emph{reciprocal}, or bidirectional. For a node in a network, let her follower (or following) set in the target network (e.g., Pnt or Lfm) be $ind$ (or $out$) and her friend set copied from the source network (e.g., Fb) be $fr$. Then the reciprocity ratios of that user in the entire target networks, and its partition into Fb-copied, and native networks are as follows:
\[R_{copied} = \frac{\vert fr \cap ind \cap out \vert}{\vert fr \cap (ind \cup out)\vert}, \]
\vspace{-2mm}
\[R_{native} = \frac{\vert(ind-fr) \cap (out-fr)\vert}{\vert(ind-fr) \cup (out-fr)\vert}. \]

On some services like Pinterest,  users \textit{follow} others unilaterally, creating directional links. We study the extent to which \textit{follow} acts are reciprocated and become bidirectional. On other services like Last.fm, users initiate a \textit{friend} request, which needs to be approved by the other party before a friendship link is instantiated, creating bidirectional links by default.
In this case, we induce a directional network by using data about historical friendship requests, treating the initial friend request as a \textit{follow}, and examine the extent to which such requests are approved by the other party, creating reified bidirectional friendship links.

\begin{figure}[!htbp]
\centering
\includegraphics[width=0.9\columnwidth]{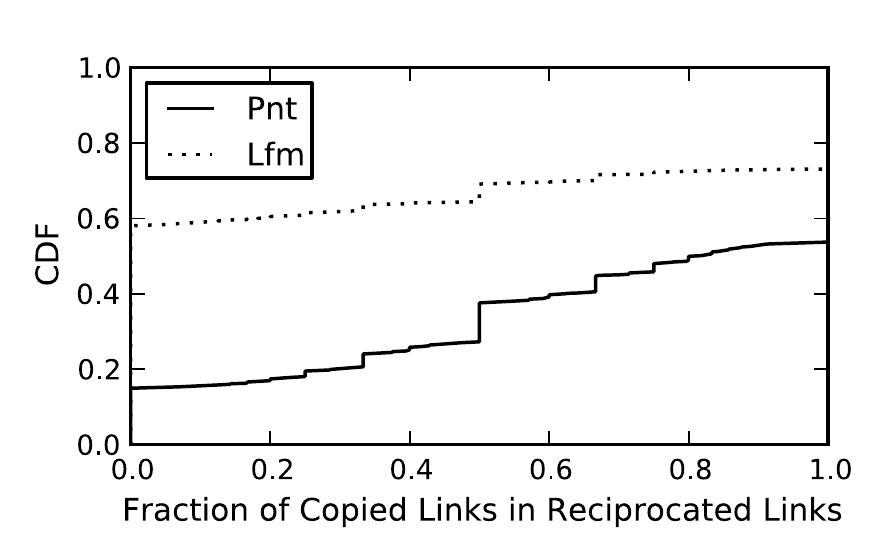}
\caption{CDF of per-user fractions of Fb-copied links among reciprocated links in target networks. Many users have high proportions of Fb-copied links implying that copied links are important for establishing bidirectional or reciprocated relationships.} 
\label{fig:fb_frac_in_reciprocated_links}
\end{figure}

\reffig{fig:reciprocity-cdf} shows that in both Pinterest and Last.fm, the reciprocity ratio is higher in links which are also found on Facebook, than on natively created links. Although in some cases, a link copied in one direction could be reciprocated by the other party merely in order to be ``social'' or ``polite'', the link creation creates an opportunity for social interaction on the target website, and reciprocity could promote positive bidirectional social interactions (We verify this in \S\ref{sec:copy-props-implications}).  The figure also indicates that  the reciprocity ratio for Last.fm is significantly higher than Pinterest. This is consistent with  user studies in previous work~\cite{changtao_zhong_sharing_2013} which found that Last.fm users easily accept requests. \reffig{fig:fb_frac_in_reciprocated_links} shows that copying is extremely important for establishing reciprocal relationships. In Pinterest, a large proportion of users' reciprocal links  are in fact those copied from Facebook. In Last.fm it is  slightly different: the fraction is relatively smaller than Pinterest, which we think is again because users tend to accept  requests easily.

\subsection{Copied network shows higher clustering}
\label{sec:copied-higher-clustering}
Next we explore the impact of copying on another popular measure of a strong social structure, clustering or the degree to which users share common friends. \reffig{fig:cc-cdf} shows that in both websites, users have much higher clustering co-efficients on the copied network  than on the network natively created on the website. Thus copying not only promotes reciprocal social interactions, but also creates a much denser social network structure in the target website.

\subsection{Copying enhances connectivity}
The increased clustering and reciprocity are properties relating to local structure around a node. Copied links are also crucial for connectivity, a global (network-wide) property. \reffig{fig:components} confirms that both the Pinterest and Last.fm copied networks have a giant component. The largest component comprises 0.91 (Pinterest) and 0.93 (Last.fm) of all the connected nodes (i.e., nodes present on both source and target networks). Furthermore, this component encompasses 0.53 (Pinterest) and 0.66 (Last.fm) of all the nodes in the corresponding target network.

\begin{figure*}[!tbp]
\centering
\hspace*{-5mm}
\subfloat[Repin network samples reciprocal links more]{
\includegraphics[width=0.72\columnwidth]{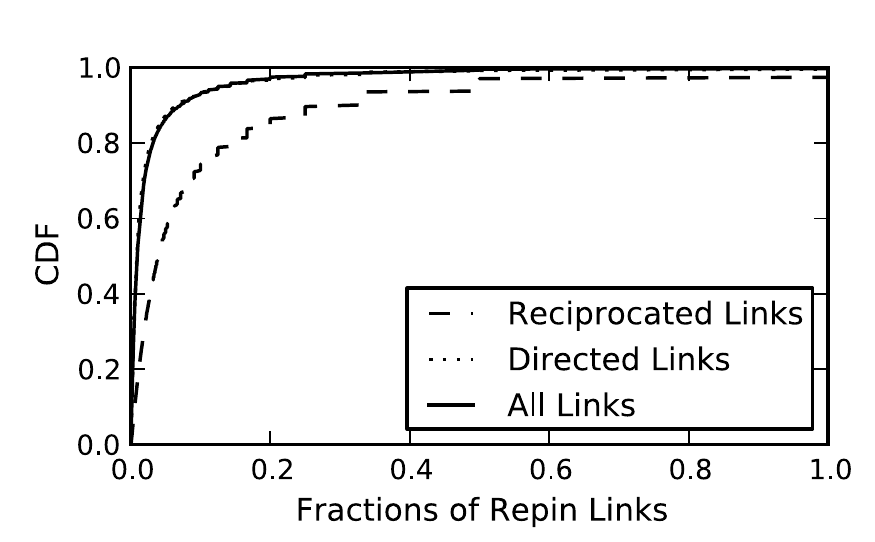}
\label{fig:cdf-repin-reciprocity}
}
\subfloat[Repin network shows higher clustering]{
\includegraphics[width=0.72\columnwidth]{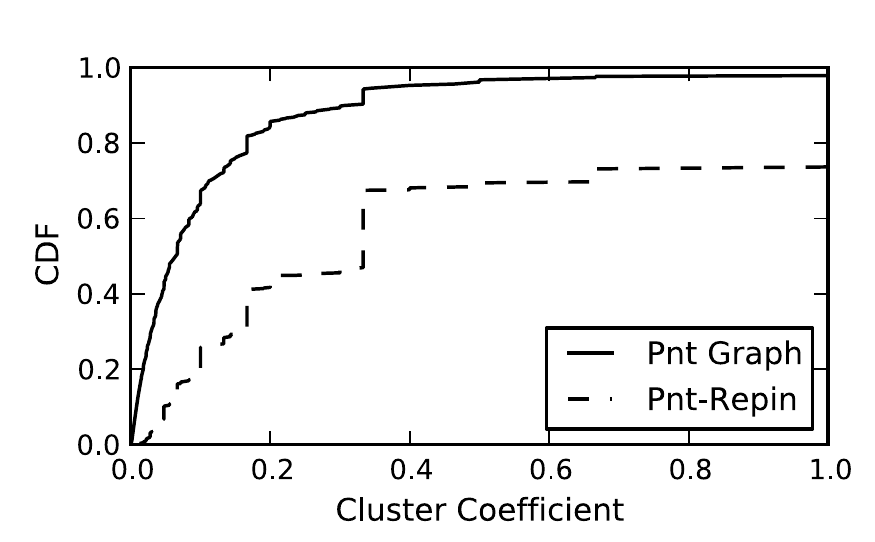}
\label{fig:social-repin-cc}
}
\subfloat[Repin network selects copied links more]{
\includegraphics[width=0.72\columnwidth]{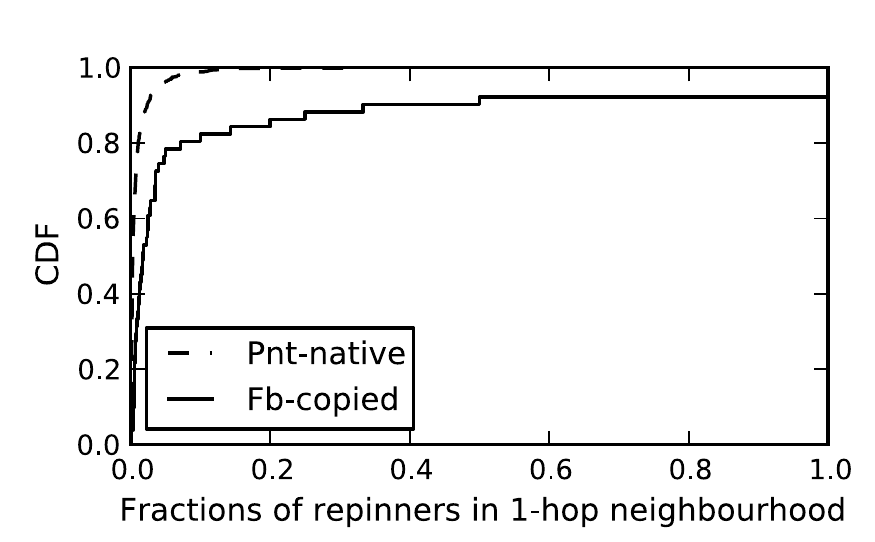}
\label{fig:action-ratio-pnt-fb}
}
\caption{How the social repin network samples the Pinterest graph (0-valued points not shown): (a)  CDF of  fraction of users' reciprocated and unreciprocated (directed) links, which are included in the repin network. A greater fraction of reciprocated links than directed links have repin activity. (b) CDF of users' clustering coefficients in the Pinterest graph and the repin network. The repin network has higher clustering, indicating that users' social repins are directed more at closer friends. (c) CDF of the fractions of users' natively created  and copied (Fb-copied) links which are sampled by the repin network. Copied links tend to have more repins.}
\end{figure*}

\subsection{Implications for social interactions}
\label{sec:copy-props-implications}
So far, we have shown that copying links results in a higher level of reciprocity and clustering,  representing a stronger and denser social structure than its  low-clustering and low-reciprocity native counterpart. While these properties are \emph{expected} to improve social interaction~\cite{Macskassy-Twitter-2012,teng_longevity_2013}, we ask whether the benefits of these structural properties are seen in the social interactions of the target network.

In order to determine the benefits of a close-knit structure, we examine one of the most popular activities on the Pinterest network, repinning. Our main question is whether copied links promote higher levels of repins. To measure this effect, we first define the concept of a \emph{social} repin, which is a repin in which a user repins a pin of someone whom she follows. We then define the \emph{social repin network}, as the subgraph of links in the Pinterest network over which at least one social repin happens in our data.

We examine how the social repin network selectively samples the underlying network of Pinterest. First, we ask what proportion of a user's reciprocated and directed (unreciprocated) links have incurred  repins. \reffig{fig:cdf-repin-reciprocity} shows that repins happen more easily over reciprocated links. Next, in \reffig{fig:social-repin-cc} we compare the clustering coefficient of users in the social repin network to the clustering coefficient of the underlying graph. Users have significantly higher clustering coefficient when we remove the links over which no repins happen. This suggests that social interactions tend to be directed towards the closer friends of a user, within highly clustered communities.

These results show  that the social repin graph is richer in reciprocated links and is more highly clustered than the underlying network. Since reciprocal links and high clustering nodes will have more social repins,  it is straightforward to infer that the copied network, which is higher in both reciprocity and clustering coefficient, should promote more social repins. This is proved by \reffig{fig:action-ratio-pnt-fb}, which shows that a larger fraction of social repinners tend to be from the copied network than from the natively created network.


\section{Weaning from Facebook}
\label{sec:weaning}

While copying links provides instant bootstrapping advantages by incurring a close-knit local structure (i.e., high reciprocity and clustering), there is a limit to which a user can copy links from Facebook. Beyond a certain point, a user may no longer find other Facebook friends to copy over. It is natural to ask whether this creates engagement bottlenecks for users as they become more prolific on the target network, or whether they find alternative solutions.

In this section, we describe a collective ``weaning'' process, through which users move away from their reliance on Facebook copied links to building new relationships natively on target websites. We find that users, as they become more active and influential on Pinterest and Last.fm, establish more native links within these services and copy less from Facebook. We discuss why users ``go native'' in this way and suggest a possible cause: through native links, users may find  others similar to themselves on the target website.

\subsection{Measures of activity and influence}

To quantify the level of user activity on Pinterest, we employ three different measures: the numbers of boards created, pins made (including repins of other users' pins), and likes of others' pins. The level of influence of a user is merely the activity of \emph{other} users directed towards that user, i.e., the number of repins and likes received by that user for her pins. In the case of Last.fm, while there is no direct measure of social influence, we have two measures of activity: the number of scrobbles and number of hits to the website.  

The results of this section are robust in the sense that they all hold for each measure of activity and influence defined above. Due to space limitations, however, results are selectively shown for only some of these measures.

\subsection{Active and influential users copy fewer links}

In order to study levels of copying, we introduce a measure called the \emph{copy ratio}. Denoting the set of all friends in the target network as $all$ and the friend set copied from the source network (i.e., Facebook) as $fr$, the copy ratio in a undirected network, such as Last.fm's, is defined as:
\[CR =  \frac{\vert all \cap fr \vert }{\vert all\vert}\]

For a directed network, representing a node's follower (resp., following) set in the target network (i.e., Pinterest) by $ind$ (resp., $out$), we define the \emph{follower copy ratio} and \emph{following copy ratio} as:
\[CR_{ind} =  \frac{\vert ind \cap fr \vert }{\vert ind\vert}\]
\[CR_{out} =  \frac{\vert out \cap fr \vert }{\vert out\vert}\]

\label{sec:activity-vs-copy}
\begin{figure}[!htbp]
\centering
\includegraphics[width=0.9\columnwidth]{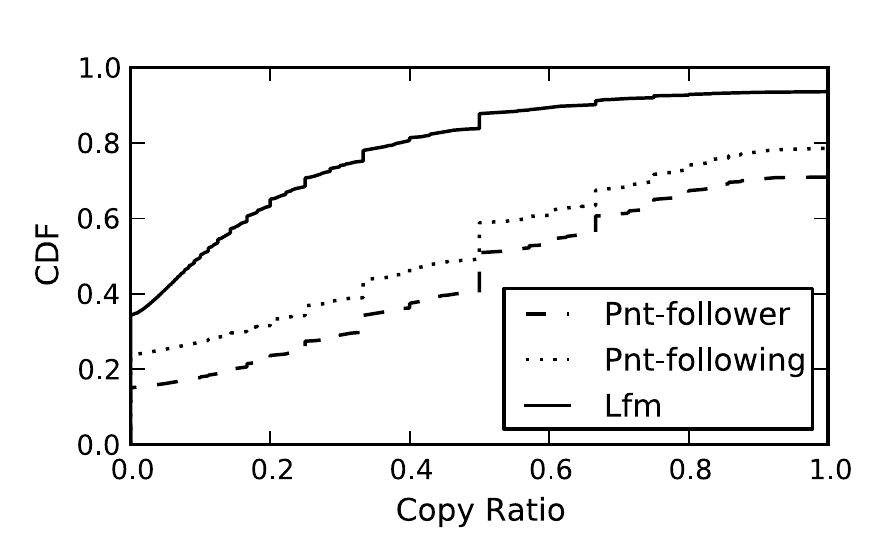}
\caption{CDF of copy ratio for connected users.}
\label{fig:copy-ratio-cdf}
\end{figure}

\begin{figure*}[!htb]
\centering
\hspace*{-5mm}
\subfloat[]{
\includegraphics[width=0.72\columnwidth]{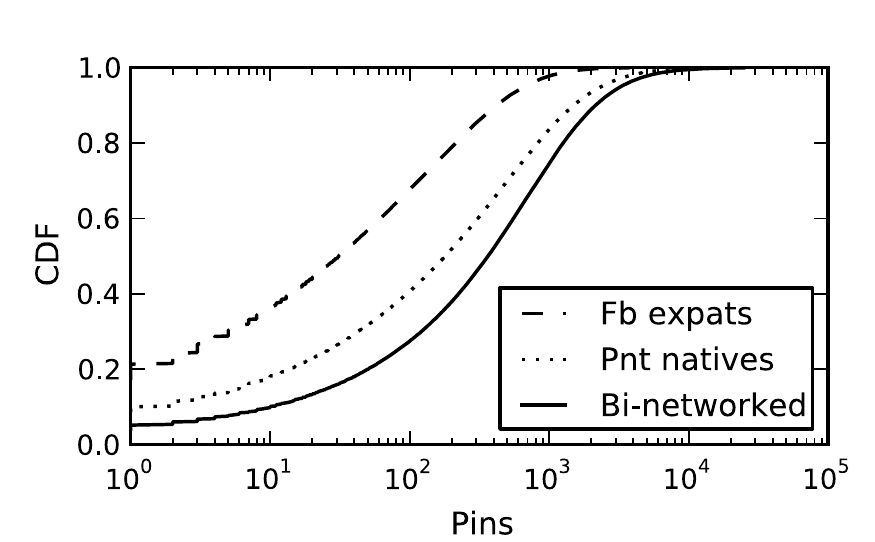}
\label{fig:copy-ratio-activities-pins}
}
\subfloat[]{
  \label{fig:copy-ratio-activities-scrobbles}
  \includegraphics[width=0.72\columnwidth]{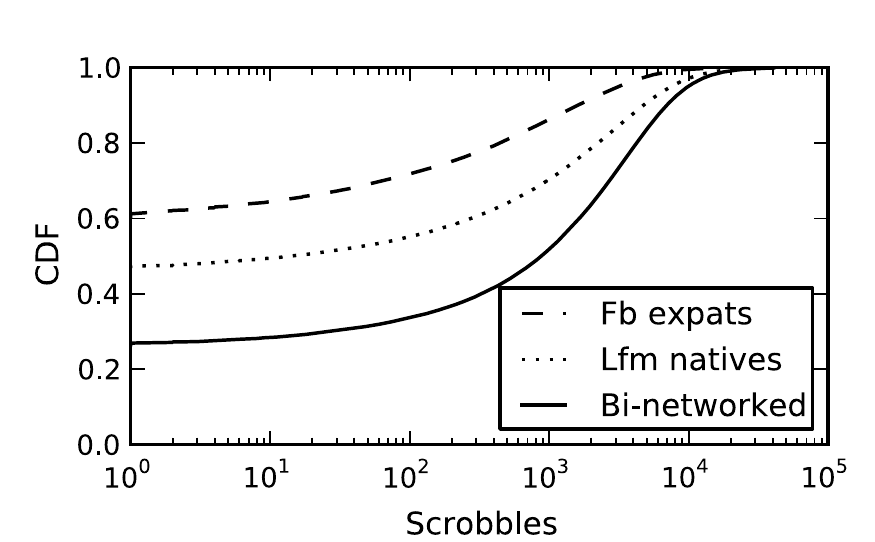}
}
\subfloat[]{
  \label{fig:copy-ratio-activities-likes}
  \includegraphics[width=0.72\columnwidth]{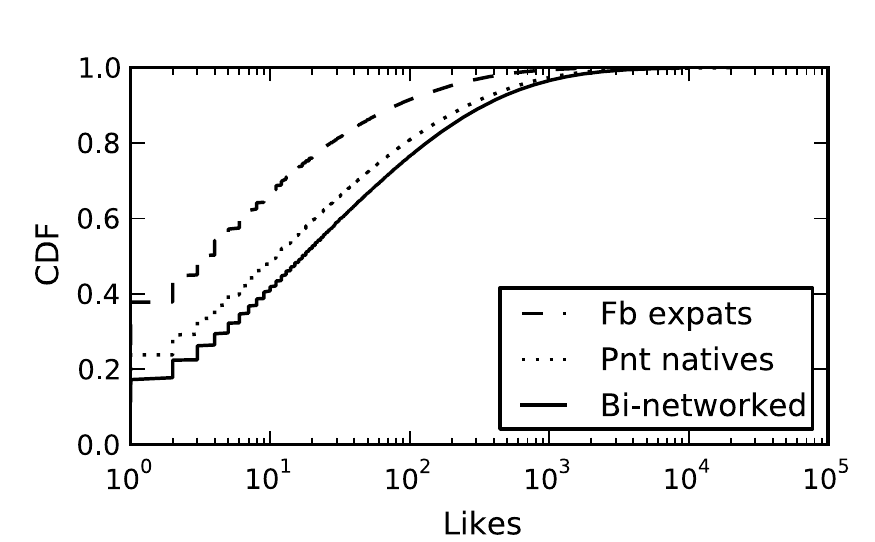}
}
\\
\vspace*{-4mm}
\hspace*{-5mm}
\subfloat[]{
\includegraphics[width=0.72\columnwidth]{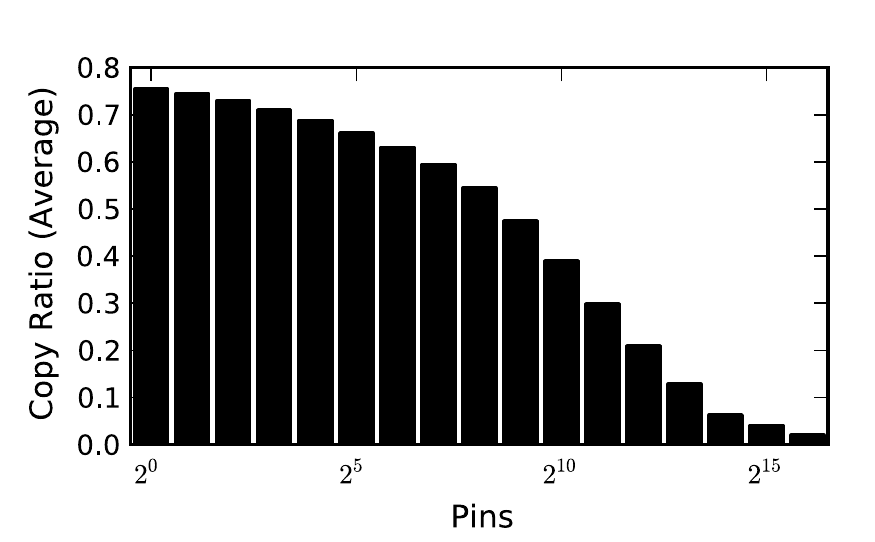}
\label{fig:active-user-copy-ratio}
}
\subfloat[]{
  \includegraphics[width=0.72\columnwidth]{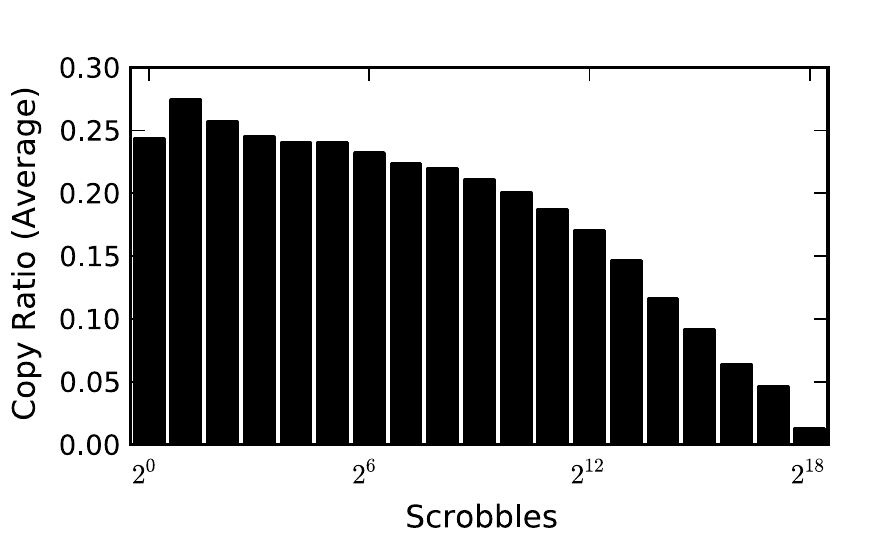}
\label{fig:active-user-copy-ratio-lf}
}
\subfloat[]{
  \includegraphics[width=0.72\columnwidth]{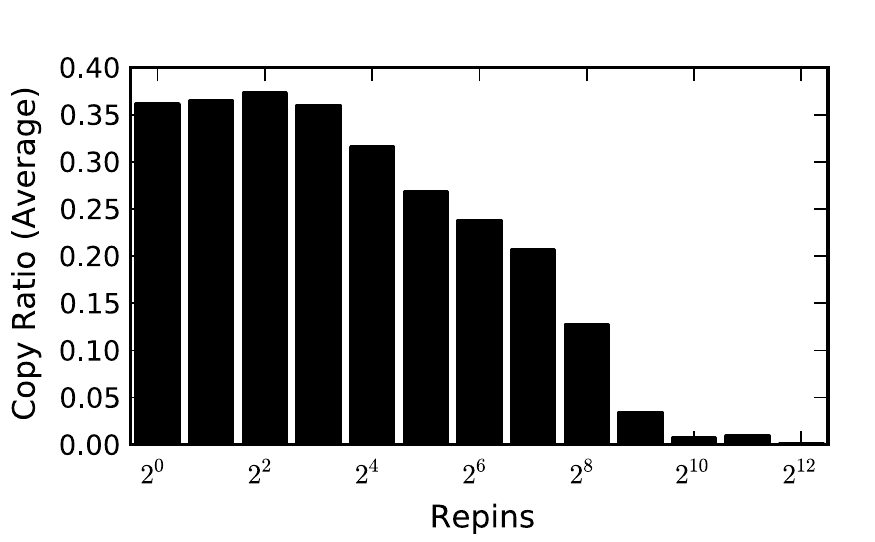}
\label{fig:active-user-copy-ratio-pins}
}
\caption{(a-c) Facebook-expats (CR=1) exhibit lesser activity in terms of pins, likes and scrobbles, compared to Pinterest and Last.fm natives (CR=0) and bi-networked users ($0<CR<1$), who are the most active. (d-e) Higher activity levels measured by pins (scrobbles) are associated with lower following copy ratio in Pinterest (Last.fm). (f) Users who are influential on Pinterest, as measured by repins, tend to have lower copy ratios.}
\end{figure*}

\reffig{fig:copy-ratio-cdf} shows the cumulative distribution of copy ratios (CR). Based on the trend shown, we may divide users into three groups. Approximately 20\% of users are \emph{Pinterest natives} (resp., \emph{Last.fm natives}), who only create links natively on the website (i.e., CR=0). A second category (20--30\%) link only to their Facebook friends (i.e., CR=1) and can be termed \emph {Facebook expats}. The majority (50--60\%), however, are \emph{bi-networked}, relying on a mixture of both native and copied links (0$<$CR$<$1). The Last.fm network contains a larger fraction of natives who have zero copy ratio (CR=0) and a smaller fraction of Facebook expats (CR=1) compared to Pinterest.

Next, we compare the activity of these three categories of users in \reffig{fig:copy-ratio-activities-pins}--c. All combinations of  activity measures and the two copy ratios show that Facebook expats (CR=1) whose social links are entirely copied from Facebook are the least active, whereas bi-networked users (0$<$CR$<$1) with a mixture of native and copied links are the most active. Pinterest and Last.fm natives (CR=0) who do not copy at all are in the mid range. This implies that users who start with the bootstrapping advantage tend to move away from   reliance on the existing Facebook network and start building new links natively (hence becoming bi-networked) as they become active members on Pinterest or Last.fm.

\reffig{fig:active-user-copy-ratio}--e drill down further and examine how the copy ratios change as activity levels increase, for the case of pins and scrobbles in Pinterest and Last.fm respectively. This demonstrates a clear inverse relationship between the activity levels and copy ratio, with users who pin or scrobble a lot tending to have lower levels of copying---that is, higher activity levels are associated with lower copy ratio. \reffig{fig:active-user-copy-ratio-pins} shows that this result extends to measures of influence on Pinterest. We find that users who are influential, measured by repins, tend to have lower copy ratios. (In the case of Last.fm, we omit this analysis due to the lack of an influence measure.) Overall, the results above indicate that as users settle down on the new service and become more active and influential, their investment in natively formed links increases proportionally.

\begin{figure*}[tb]
\centering
\hspace*{-5mm}
\subfloat[]{
  \label{fig:active-user-social-ratio-pins}
  \includegraphics[width=0.72\columnwidth]{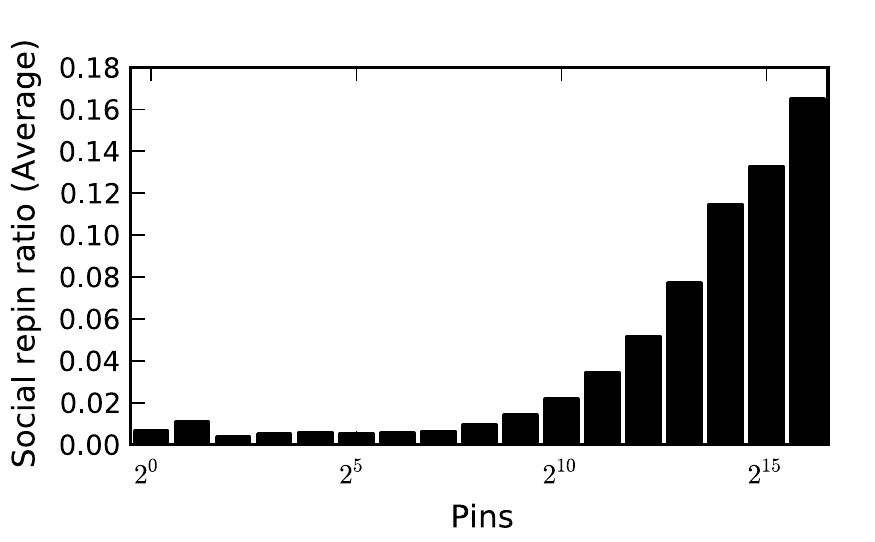}
}
\subfloat[]{
  \label{fig:active-user-social-ratio-likes}
  \includegraphics[width=0.72\columnwidth]{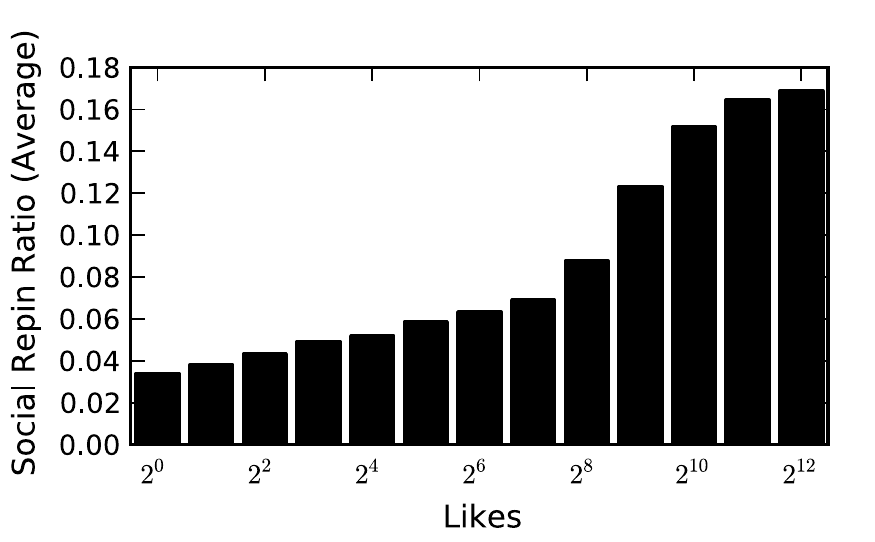}
}
\subfloat[]{
  \label{fig:social-repin-ratio-activities}
  \includegraphics[width=0.72\columnwidth]{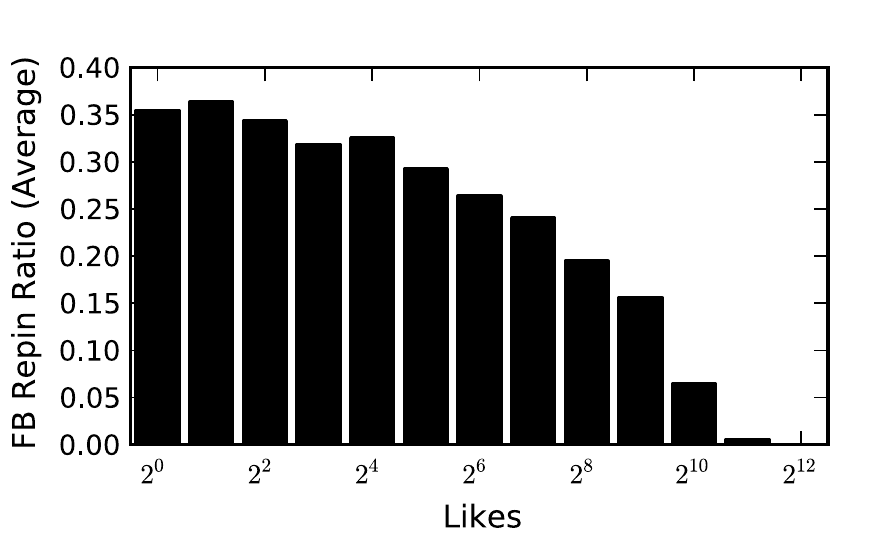}
}
\vspace*{-4mm}
\caption{(a) Social repins  increase proportionally as activity level, measured by pins, increases.  (b) The same increasing trend is shown for influence, measured by likes received.  (c) However, the fraction of social repins over links \emph{copied} from Facebook decreases proportionally as influence, measured by likes received, increases.}
\vspace*{-5mm}
\end{figure*}

\subsection{Influential and active users remain social, but with native rather than copied friends}

Next we take a deeper look at the relationship between the increase in activity or influence level of a user and his level of social interaction. In order to quantify the level of social interaction, we again define the concept of a \emph{social repin} as a repin where the user who is repinning follows the original pinner. We define a user's \emph{social repin ratio} for activity (or influence) as the fraction of social repins made (or  received) among all repins made (or received).

\reffig{fig:active-user-social-ratio-pins}--b shows that users who are more active (or influential) tend to make (or  receive) proportionally more  social repins in relation to their activity (or influence) level, \emph{confirming that social interaction continues to be increasingly essential} for active (or influential) users.

We also focus on social repins and ask whether copied links promote social repins. We define   the \emph{Facebook repin ratio} for activity (or influence) as the fraction of social repins made (or received) over Fb-copied links among all social repins made (or received). \reffig{fig:social-repin-ratio-activities} reveals that as activity (or  influence) levels increase, social repins happening over copied links decrease.

\begin{figure*}[htbp]
\centering
\hspace*{-5mm}
\subfloat[Native friends have similar interests]{
\includegraphics[width=0.72\columnwidth]{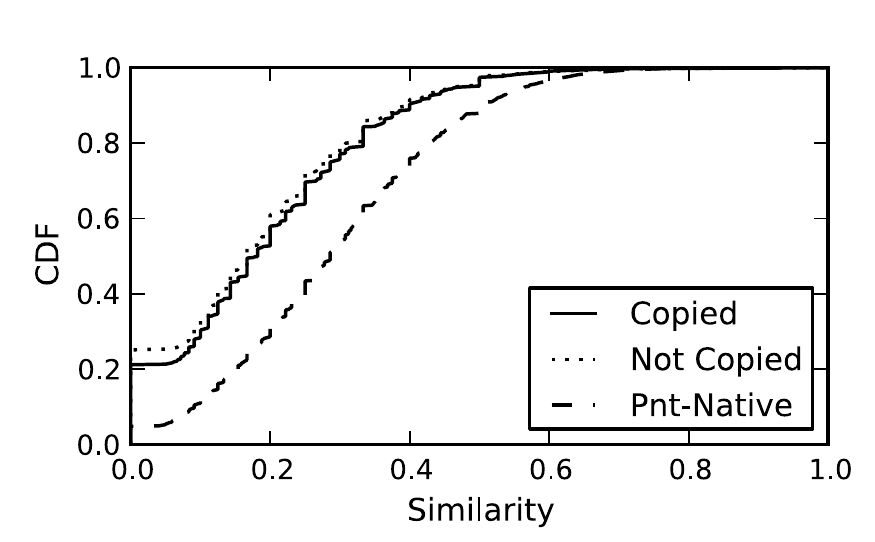}
\label{fig:board-sim-cdf}
}
\subfloat[Close friends are preferentially copied]{
  \label{fig:bias-close-friend}
  \includegraphics[width=0.72\columnwidth]{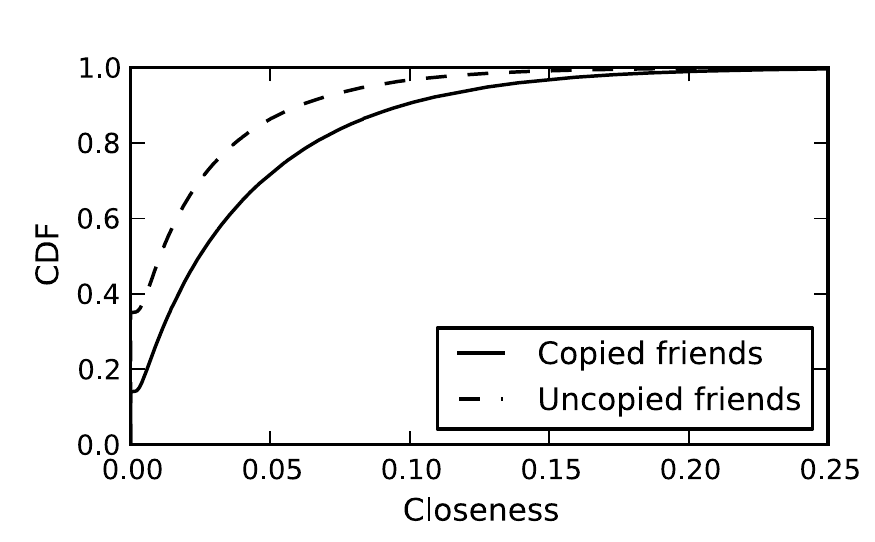}
}
\subfloat[Native friends are FoFs of copied friends]{
\includegraphics[width=0.72\columnwidth]{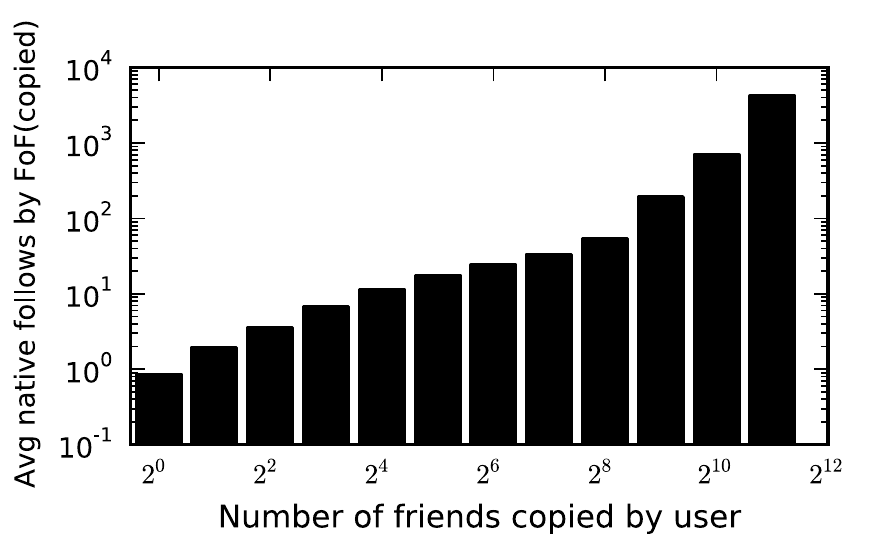}
\label{fig:copied-native-clustering}
}
\caption{User choice in copying links (Pinterest data) (a) CDF of similarity between users linked by copied, uncopied and pnt-native links, showing that native friends are more similar to a user than copied friends, but copied and uncopied friends do not differ significantly in tastes. (b) Per-user CDF of closeness between copied and uncopied friends. It shows that copied friends are closer than uncopied friends. (c) The more friends a user copies and follows in the target network, the more follows she gets, from exposure to friends of the copied friends.}
\end{figure*}

\subsection{Weaning, biases and community evolution}
\label{sec:weaning-impl}
We conclude by asking how the nature of the target network community would evolve as users `wean' from copying to make more native links. To understand this, we study user preferences or biases in the kinds of links they copy and the links they make natively. We also seek to understand the role that copying plays in creating more native links.

User studies in previous work on Pinterest~\cite{changtao_zhong_sharing_2013} identified that Pinterest users most value the social aspect of the service that helps them find people with similar tastes in pictures. Therefore, we examine whether natively created links on Pinterest enables discovery of individuals with a more similar taste than those with copied links. Specifically, if $I_1$ is the set of user $u_1$'s board categories, and $I_2$ is the set of $u_2$'s, we define their similarity as 
$s = \frac{\vert I_1 \cap I_2\vert}{\vert I_1 \cup I_2\vert}.$
\reffig{fig:board-sim-cdf} confirms that according to this measure, users connected by native Pinterest links are more similar to each other than those connected by Facebook-copied links. 

\reffig{fig:board-sim-cdf} also shows that there is no difference in similarity between users who are copied and users who are not copied over from Facebook. This implies that users are not selecting Facebook friends to copy based on similarity. On Last.fm, native links also show higher similarity than Facebook copied links, which in turn show higher similarity than Facebook links which were deliberately not copied. We conjecture that on some websites like Pinterest, there might be social norms at play which are being used to decide which links could be copied over.

In \reffig{fig:bias-close-friend}, we study whether closeness of friends has a role in deciding which friends to copy. 
In our analysis, we use the similarity of users' friend lists to show their closeness: if $A$'s friend list is $L_A$ and $B$'s is $L_B$, we say their closeness is $\frac{L_A \cap L_B}{L_A \cup L_B}$. \reffig{fig:bias-close-friend} shows  that closeness between copied friends is higher than between uncopied friends. In Last.fm, however, there is no significant difference between closeness of copied and uncopied friends.

Together these results suggest that Pinterest users tend to use the ``friend finder'' tool to copy close friends they know from established source networks like Facebook, but when they discover new friends on the target network, they tend to prefer users with similar tastes. Thus, as native links become more important and numerous than copied links, we expect the target network to become more interest-based. By contrast, Last.fm users tend to prefer copying and natively linking to users who share similar music tastes. Thus, while there is no universal pattern for how users on different target networks copy links, it appears that in both cases, the links, and hence the target communities, tend to become more interest-based over time.

However, copying continues to be important for  the creation of native links over which interaction happens, even in networks like Pinterest, where copying appears to be governed by norms of social closeness -- \reffig{fig:copied-native-clustering} examines links over which social repin interactions happen over a sample representative day, and shows that users who have copied more of their friends from the source to target network, tend to have more native followers who are \emph{friends of her friends} on the target network. i.e., copying creates the opportunity for users in the immediate social community of nodes in the copied sub-graph to discover and follow them, creating new native links, over which social interaction happens.


\section{Related Work}

Many websites try to incorporate a social networking aspect to enhance user engagement and community interaction.  Social networks are known to facilitate the formation of learning communities, foster student engagement and reflection, and enhance overall user experience for students in synchronous and asynchronous learning environments~\cite{baird_neomillennial_2005}.  Social networks are also the core of the design of new user-driven communities around health issues~\cite{eysenbach_medicine_2008} and have been utilised to facilitate community formation in environments ranging from professional settings~\cite{lee_experiments_2013} to online games~\cite{choi_why_2004,ducheneaut_social_2004}. Including the above studies, most existing research on the formation and evolution of online social network communities has focused on single networks. In contrast, this paper evaluated interactions between two \emph{different} networks: a generic social network  and a target  network on the content-driven website.

Multilayer networks (or also called multiplex, heterogeneous, interdependent, or  multi-relational networks in literature) describe the fact that users may belong to different social networks (or layers) at the same time in real world. Each network layer could have particular features different from the others. A number of studies have looked into multilayer networks, including modelling of the formation and evolution of multilayer networks based on preferential attachment models~\cite{Magnani-formationMultiplex-2013, Nicosia-model-2013, Podobnik-multiplex-2012}.  In particular,  resilience of cooperative behaviours is known to enhanced by a multilayer structure~\cite{Gomez-Coopertion-2012} and in some cases, cascading failures may occur in interacting networks~\cite{buldyrev_catastrophic_2010}. Multilayer structures are also known to speed up diffusion in networks~\cite{Gomez-diffusion-2013}.

We proposed a copying process for networks as a model called Link Bootstrapping Sampling (LBS). This model can be seen as a variation of Induced Subgraph Sampling (ISS)~\cite{Kolaczyk_network_2009}, which  randomly selects a subset of nodes and observes all links between selected nodes.  Compared to ISS, the LBS model introduces an additional link sampling step: each selected node further selects a \emph{subset} of its links for observation. Then, based on this model, we derived conditions for the emergence of a giant connected component in copied networks.  Lee \textsl{et al.}~\cite{lee_correlated_2012} also examine the emergence of a giant connected component in multi-layer networks. However their study is restricted to the Erd{\H o}s-R{\'e}nyi model and seeks to answer a different question of how the correlation of node degrees in different layers affects the emergence of  a giant component.

Another aspect of this paper was the large-scale empirical study across two different networks. Empirical analysis of multiple networks is relatively uncommon.  Szell~\cite{szell_multirelational_2010} collected data from an online game and extracted  networks of six different types of one-to-one interactions between the players. Then, both reciprocity and clustering were studied for each layer of the network. In contrast, our dataset shows the process of copying links between two independent websites, the source and target, where the original purpose of the link in the source network may be quite different from the intended purpose for the copied link in the target network.

Finally, this paper is related to the series of studies that investigate the motivation of users in creating social network links. One study found that professionals use internal social networking to build stronger bonds with their weak ties and to reach out to employees they do not know~\cite{dimicco_motivations_2008}. Another study identified that social links have high predictive power in determining which newcomers will continue to engage with the service in the future~\cite{burke_feed_2009}. Other studies have demonstrated that properties such as reciprocity and clustering promote  interaction in natively created social graphs~\cite{Macskassy-Twitter-2012,teng_longevity_2013}. We have shown that such positive effects of social links also apply to links copied from unrelated \emph{external} social networks.


\section{Conclusions}

This paper studied the impact of social bootstrapping---the act of copying one's social ties or links from a source network to a target network. This is a popular practice enabled by many new social network services and has implications on how a new online social network community can grow quickly. We gathered massive amounts of data from Facebook, Pinterest, and Last.fm involving tens of millions of nodes and billions of links to understand this new phenomenon. We proposed a simple analytical model and used it to gain insight into the social bootstrapping process. Among a number of findings, we highlight that a ``copying'' process is useful to initiate social interaction in the target network, as one may expect. However, a ``weaning'' process, where a user moves away from  copied social links and builds social relationships natively in the new network is essential for longer lasting user engagement. To the best of our knowledge, this paper is the first to utilize large-scale cross network data in understanding the interplay between heterogeneous services in terms of bootstrapping a network and engaging users to form a cohesive, interacting community. 
%
\section*{Acknowledgements}
This research was partially supported by the UK Engineering and
Physical Sciences Research Council via Grant No.\ EP/K024914/1 and the Basic Science Research Program through the National Research Foundation of Korea funded by the Ministry of Science, ICT \& Future Planning (2011-0012988).

\balance
\bibliographystyle{abbrv}
\bibliography{bib/pnt-fb}
\end{document}